\documentclass[aps,prd,reprint,twocolumn,superscriptaddress,showpacs]{revtex4-2}

\usepackage{graphicx}
\usepackage{mathrsfs}
\usepackage{bm}
\usepackage{amsmath}
\usepackage{dcolumn}
\usepackage{epstopdf}
\usepackage{dsfont}
\usepackage{amssymb}
\usepackage{tabularx}
\usepackage{array}
\usepackage{float}
\usepackage{color}
\usepackage{epstopdf}
\usepackage{mathrsfs}
\usepackage[colorlinks, linkcolor=blue,anchorcolor=blue,citecolor=blue,urlcolor=blue]{hyperref}
\usepackage{multirow}

\begin{document}

\title{Valley- and Spin-Dependent Electronic and Transport Properties of Two-Dimensional Altermagnetic Titanium-Based Chalcogenide Halides}
	
\author{Ruo-Yu Ning}
\affiliation{School of Physics, Northwest University, Xi'an 710127, China}

\author{Zhi-Hua Yan}
\affiliation{School of Physics, Northwest University, Xi'an 710127, China}

\author{Jin-Yang Li}
\affiliation{School of Physics, Northwest University, Xi'an 710127, China}

\author{Yong-Kun Wang}
\affiliation{School of Physics, Northwest University, Xi'an 710127, China}
	
\author{Si Li}
\email{sili@nwu.edu.cn}
\affiliation{School of Physics, Northwest University, Xi'an 710127, China}
\affiliation{Shaanxi Key Laboratory for Theoretical Physics Frontiers, Xi'an 710127, China}
\affiliation{Peng Huanwu Center for Fundamental Theory, Xi'an 710127, China}
\affiliation{Fundamental Discipline Research Center for Quantum Science and Technology of Shaanxi Province, Xi'an 710127, China}
	
\begin{abstract}
Altermagnets (AMs) combine fully compensated magnetization with momentum-dependent spin splitting, yet intrinsic altermagnetic materials exhibiting exceptional valley characteristics remain scarce. Here, we identify monolayer titanium-based chalcogenide halides, Ti$_2X_2Y$ ($X$ = F, Cl, Br, I; $Y$ = O, S, Se, Te), as a new family of two-dimensional (2D) altermagnetic valley materials. These monolayers exhibit robust $d$-wave altermagnetic order, semiconducting band gaps, and pronounced spin-polarized valley characteristics. We show that uniaxial strain breaks the valley degeneracy, inducing giant valley splitting together with a tunable piezomagnetic response. An in-plane electric field generates noncollinear spin currents, while spin--orbit coupling gives rise to the anomalous Hall effect, valley-selective linear dichroism, and the magneto-optical Kerr effect. These findings establish Ti$_2X_2Y$ monolayers as a versatile platform for exploring spin- and valley-dependent electronic, optical, and transport phenomena in 2D altermagnets.

\end{abstract}
	
\maketitle
\section{Introduction}
Altermagnets (AMs) have recently emerged as a fascinating magnetic phase in condensed matter physics, uniquely bridging key characteristics of ferromagnets and antiferromagnets~\cite{smejkal2022conventional,smejkal2022emerging,wu2007fermi,hayami2019momentum,yuan2020giant,ma2021multifunctional,liu2022spin,bai2024altermagnetism,fender2025altermagnetism,song2025altermagnets}. Although they exhibit a fully compensated macroscopic magnetization in real space, similar to conventional antiferromagnets, their opposite spin sublattices are related by specific crystal rotations or mirror symmetries rather than by spatial inversion or translation~\cite{smejkal2022conventional,smejkal2022emerging,liu2022spin,xiao2024spin,chen2024enumeration,jiang2024enumeration}. This unconventional symmetry is crucial, as it intrinsically lifts spin degeneracy even in the absence of spin–orbit coupling (SOC), giving rise to robust momentum-dependent spin splitting across the entire Brillouin zone (BZ). As a result, AMs host a variety of exotic phenomena, including the anomalous Hall effect~\cite{vsmejkal2020crystal,feng2022anomalous,wang2025symmetry}, symmetry-driven spin-current generation~\cite{wu2007fermi,ma2021multifunctional,gonzalezhernandez2021efficient,bose2022tilted}, giant tunneling magnetoresistance~\cite{shao2021spin,smejkal2022giant}, spin Seebeck and Nernst effects~\cite{cui2023efficient}, as well as topological states~\cite{han2024cornertronics,antonenko2025mirror,wang2025pentagonal,wang2026two,wang2025real}. Moreover, the altermagnetic state enables unconventional Andreev reflection~\cite{papaj2023andreev,sun2023andreev}, finite-momentum Cooper pairing~\cite{zhang2024finite,hong2025unconventional,sim2025pair,chakraborty2024zero}, and emergent topological superconductivity~\cite{li2023majorana,ghorashi2024altermagnetic,li2024creation,zhu2023topological}, along with ferroelectric and antiferroelectric responses~\cite{gu2025ferroelectric,duan2025antiferroelectric,zhu2025two,vsmejkal2024altermagnetic,urru2025g}. Motivated by these intriguing properties, a rapidly growing family of AM candidates has been proposed theoretically and confirmed experimentally, including representative materials such as RuO$_2$~\cite{berlijn2017itinerant,zhu2019anomalous}, MnTe~\cite{gonzalez2023spontaneous,krempasky2024altermagnetic}, CrSb~\cite{li2025topological,lu2025signature,ding2024large,zhou2025manipulation,yang2025three}, Rb$_{1-\delta}$V$_2$Te$_2$O~\cite{zhang2025crystal}, and KV$_2$Se$_2$O~\cite{jiang2025metallic}.

Beyond their unconventional spin characteristics, AMs are also driving a paradigm shift in valleytronics~\cite{ma2021multifunctional,gunawan2006valley,xiao2007valley,rycerz2007valley,yao2008valley,xiao2012coupled,cai2013magnetic,schaibley2016valley,vitale2018valleytronics,guo2024valley,li2024strain,Fan2025}. In prototypical two-dimensional (2D) valleytronic systems, such as graphene and transition metal dichalcogenides, the valley degeneracy at the K and K$'$ points is strictly protected by time-reversal ($\mathcal{T}$) symmetry. Consequently, achieving valley polarization in these systems typically requires external $\mathcal{T}$-symmetry-breaking perturbations, such as magnetic fields~\cite{cai2013magnetic,aivazian2015magnetic,srivastava2015valley,macneill2015breaking,qi2015giant,jiang2017zeeman} or circularly polarized optical excitation~\cite{mak2012control,zeng2012valley,cao2012valley,hsu2015optically,mak2018light}. In contrast, altermagnetism provides an elegant route to overcome this limitation. In AMs, valley degeneracy is governed by crystalline symmetries rather than $\mathcal{T}$ symmetry, enabling versatile, time-reversal-invariant strategies for valley manipulation, such as via mechanical strain~\cite{ma2021multifunctional,li2024strain,zhu2023multipiezo} or gate-tunable electric fields~\cite{zhang2024predictable}.
Nevertheless, the limited availability of experimentally realizable 2D altermagnetic materials with pronounced valley characteristics remains a major obstacle to their practical application. Therefore, discovering new 2D altermagnets with significant valley features and exploring their valley- and spin-dependent physical properties are of critical importance.

In this work, based on first-principles calculations and theoretical analysis, we predict a series of monolayer titanium-based chalcogenide halides, Ti$_2X_2Y$ ($X$ = F, Cl, Br, I and $Y$ = O, S, Se, Te), as altermagnetic materials with pronounced valley characteristics. These materials possess excellent stability and robust $d$-wave altermagnetism with intrinsic momentum-dependent spin splitting in the absence of SOC.
We find that these systems exhibit semiconducting behavior, with energetically degenerate valleys located at the time-reversal-invariant momenta X and Y points. Notably, the valleys carry opposite spin polarization, dictated by intrinsic altermagnetic coupling and the $\{C_{2}\parallel C_{4z}^+\}$ symmetry. Breaking this symmetry via external uniaxial strain lifts the valley degeneracy, leading to giant valley splitting and inducing a piezomagnetic response under hole doping.
Furthermore, we investigate the generation of noncollinear spin currents driven by an in-plane electric field. When SOC is included, these systems exhibit rich physical phenomena, including anomalous Hall effect, linear dichroism and magneto-optical Kerr effect (MOKE).
Our results reveal a rich landscape of valley- and spin-dependent electronic and transport properties in monolayer altermagnetic titanium-based chalcogenide halides, highlighting their promising potential for future applications in valleytronics, spintronics, and nanoelectronics.

\section{First-principles Methods}
First-principles computations were performed within the density functional theory (DFT) framework using the Vienna \textit{ab initio} Simulation Package (VASP) \cite{kresse1994ab,kresse1996efficient}, with the exchange-correlation functional described by the Perdew–Burke–Ernzerhof (PBE) parametrization of the generalized gradient approximation (GGA) \cite{perdew1996generalized}. The wave functions were expanded in a plane-wave basis set with a kinetic energy cutoff of 500 eV, and the BZ integration was performed over a $\Gamma$-centered $13 \times 13 \times 1$ $k$-mesh. Energy and force convergence criteria were strictly maintained at $10^{-7}$ eV and 0.01 eV/Å, respectively. A generous vacuum guard of 20 Å was applied along the $z$ axis to avoid artificial mirror interactions. In line with previous studies, the localized Ti $3d$ electrons were treated using the DFT+$U$ scheme with an effective Hubbard parameter of $U = 4.3$ eV~\cite{moore2024high}. We also tested $U$ values of 3 and 5 eV, as shown in the Supplemental Material (SM)~\cite{SM}.
The phonon spectra were derived from density functional perturbation theory (DFPT) calculations using a $2 \times 2 \times 1$ supercell via the PHONOPY package \cite{togo2015first}. The same supercell size was adopted for \textit{ab initio} molecular dynamics (AIMD) simulations to verify thermodynamic stability. Utilizing the Wannier90 code \cite{marzari1997maximally,souza2001maximally,pizzi2014boltzwann}, spin-resolved charge conductivities were evaluated via Boltzmann transport theory within the constant relaxation time approximation. These transport coefficients were sampled at 300 K using a highly dense $300 \times 300 \times 1$ $k$-mesh and a relaxation time of 10 fs. Crucially, to remove the vacuum artifact and align the dimensions with standard 3D bulk definitions, the transport properties were scaled by $L_z/d_{\mathrm{eff}}$, where $L_z$ and $d_{\mathrm{eff}}$ denote the total vertical lattice parameter and the effective monolayer thickness, respectively. Additionally, the Berry curvature was calculated using the VASPBERRY code \cite{kim2022circular}.

\section{RESULTS}

\subsection{Crystal structure and magnetism}
\begin{figure}[htb]
	\includegraphics[width=8.5cm]{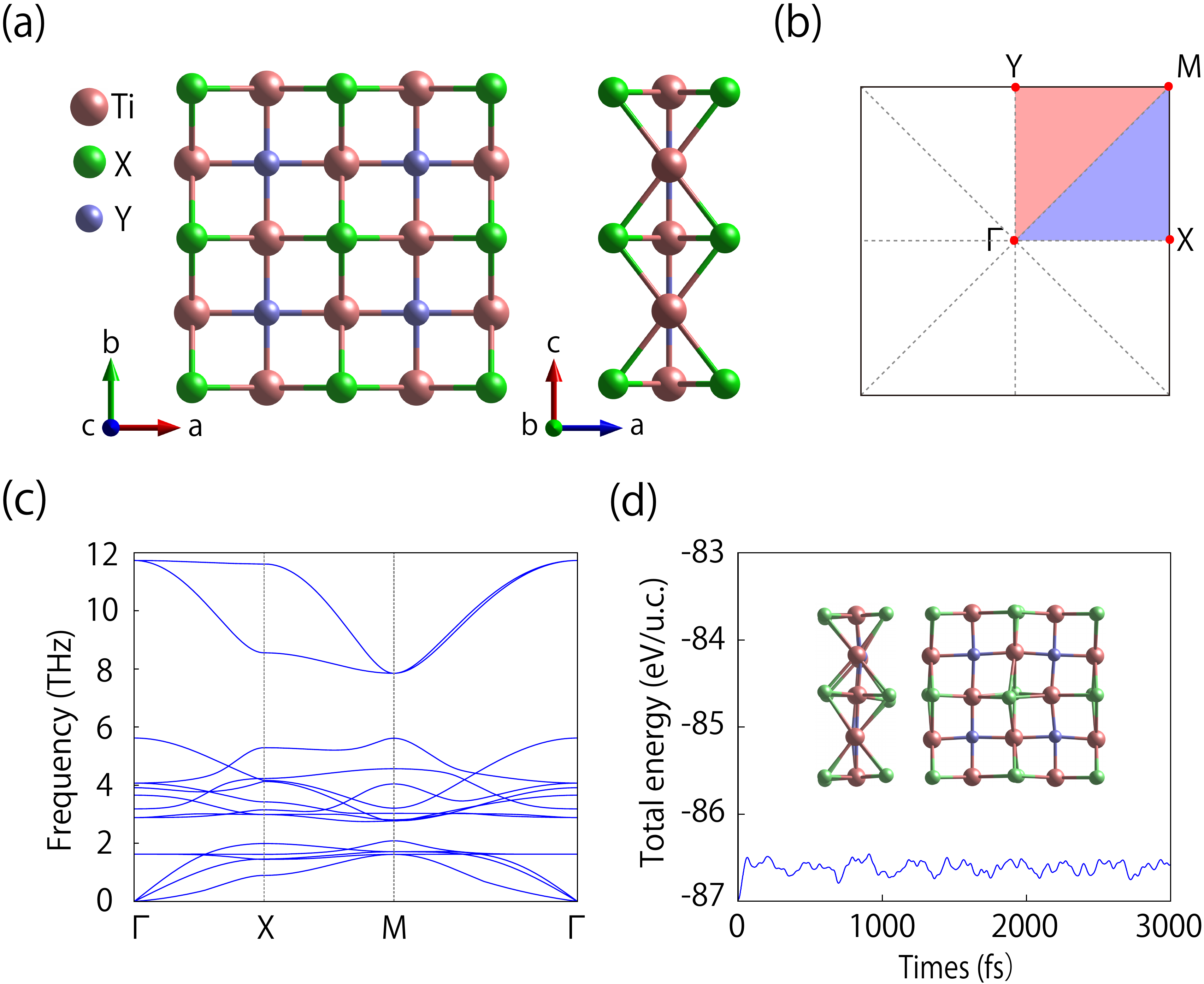}
	\caption{(a) Top and side views of the crystal structure of the monolayer Ti$_2X_2Y$ ($X$ = F, Cl, Br, I and $Y$ = O, S, Se, Te). (b) BZ with indicated high-symmetry points. (c) Calculated phonon spectrum of Ti$_2$I$_2$S. (d) Ab initio molecular dynamics (AIMD) results of Ti$_2$I$_2$S.}
	\label{fig1}
\end{figure}
The Ti$_2X_2Y$ ($X$ = F, Cl, Br, I and $Y$ = O, S, Se, Te) monolayers adopt a tetragonal lattice structure belonging to the space group $P4/mmm$ (No.~123), characterized by symmetry operations generated by $S_{4z}^+$, $C_{2x}$, and inversion $\mathcal{P}$. As illustrated in Fig.~\ref{fig1}(a), the monolayer exhibits a trilayer architecture, in which the Ti and chalcogen atoms in the middle layer are sandwiched between halogen atoms in the top and bottom layers. The corresponding BZ is shown in Fig.~\ref{fig1}(b). The optimized lattice constants for Ti$_2X_2Y$ are summarized in Table~\ref{table1}.
Given the structural and electronic similarities among these compounds, we primarily focus on Ti$_2$I$_2$S in the main text, while the results for Ti$_2$I$_2$Se, Ti$_2$I$_2$Te, as well as the electronic band structures of other candidate materials, are provided in the SM~\cite{SM}.

\begin{table*}[htb]
	\caption{Calculated properties of monolayer Ti$_2X_2Y$ ($X$ = F, Cl, Br, I; $Y$ = O, S, Se, Te), including the optimized lattice constant $a$ (\AA), energy differences between the AM and FM/SAFM/ZAFM states (eV per primitive cell), nearest-neighbor exchange parameter $J_1$ (meV), global band gap $E_g$ (eV) without SOC, and magnetic anisotropy energies (MAEs) with SOC included, defined as the energy differences between the [001] and [100] directions ($\Delta E_{\mathrm{001-100}}$) or between the [001] and [110] directions ($\Delta E_{\mathrm{001-110}}$), both in $\mu$eV per primitive cell.}
	\begin{ruledtabular}
	\begin{tabular}{ccccccccc}
		Systems  & $a$ & $\Delta E_{\mathrm{AM-FM}}$  & $\Delta E_{\mathrm{AM-SAFM}}$  &  $\Delta E_{\mathrm{AM-ZAFM}}$  & $J_1$ & $E_g$ & $\Delta E_{\mathrm{001-100}}$ & $\Delta E_{\mathrm{001-110}}$ \\
		\hline
		Ti$_2$F$_2$O   & 4.139    & $-4.09228$   & $-0.72822$  & $-1.40357$  & $-127.884$ & $1.183$ & $13.800$ & $13.810$  \\ 
		Ti$_2$Cl$_2$O   & 4.240   & $-3.86752$   & $-0.89845$  & $-0.89114$  & $-120.860$ &  $1.750$ & $34.620$ & $34.630$ \\
		Ti$_2$Cl$_2$S   & 4.870   & $-2.56726$   & $-1.63309$  & $-2.12063$  & $-80.227$ & $1.564$ & $13.800$ & $13.800$  \\
		Ti$_2$Cl$_2$Se   & 5.069   & $-2.48456$   & $-1.34685$  & $-0.56384$  & $-77.643$ & $1.750$ & $0.060$ & $0.130$ \\
		Ti$_2$Br$_2$O   & 4.281    & $-3.68930$   & $-0.83776$  & $-0.85598$  & $-115.291$ & $1.720$ & $51.590$ & $51.200$  \\
		Ti$_2$Br$_2$S   & 4.923   & $-2.40097$   & $-4.05325$  & $-0.67862$  & $-75.030$ & $1.583$ & $12.080$ & $11.720$  \\
		Ti$_2$Br$_2$Se  & 5.116   & $-2.33044$   & $-4.32767$  & $-0.598112$  & $-72.826$ & $1.752$ & $-0.590$ & $-0.440$ \\
		Ti$_2$I$_2$O   & 4.313    & $-3.48525$   & $-0.71565$  & $-0.79510$  & $-108.914$ & $1.602$ & $39.000$ & $37.890$  \\
		Ti$_2$I$_2$S   & 4.976    & $-2.11221$  & $-3.89997$ &  $-0.67381$   & $-66.006$ & $1.515$ & $-71.880$    & $-72.680$  \\
		Ti$_2$I$_2$Se  & 5.175    & $-2.08919$  & $-4.17391$ &  $-0.16623$   & $-65.287$ & $1.695$ & $-84.920$    & $-86.110$  \\
		Ti$_2$I$_2$Te  & 5.498    & $-2.03177$  & $-1.19721$ &  $-0.48797$   & $-63.493$ & $1.839$ & $-153.880$   & $-152.250$   \\
	\end{tabular}\label{table1}
\end{ruledtabular}
\end{table*}

We first examine the stability of monolayer Ti$_2$I$_2$S. The phonon dispersion, shown in Fig.~\ref{fig1}(c), exhibits no imaginary frequencies across the entire BZ, confirming its dynamical stability. To further assess thermal stability, which is essential for practical applications, we performed ab initio molecular dynamics (AIMD) simulations using a $2 \times 2 \times 1$ supercell. As shown in Fig.~\ref{fig1}(d), the structure exhibits only minor thermal fluctuations after 3000 fs at 300 K, with no bond breaking or structural reconstruction, demonstrating robust thermal stability under ambient conditions.

\begin{figure}[htb]
	\includegraphics[width=8.5cm]{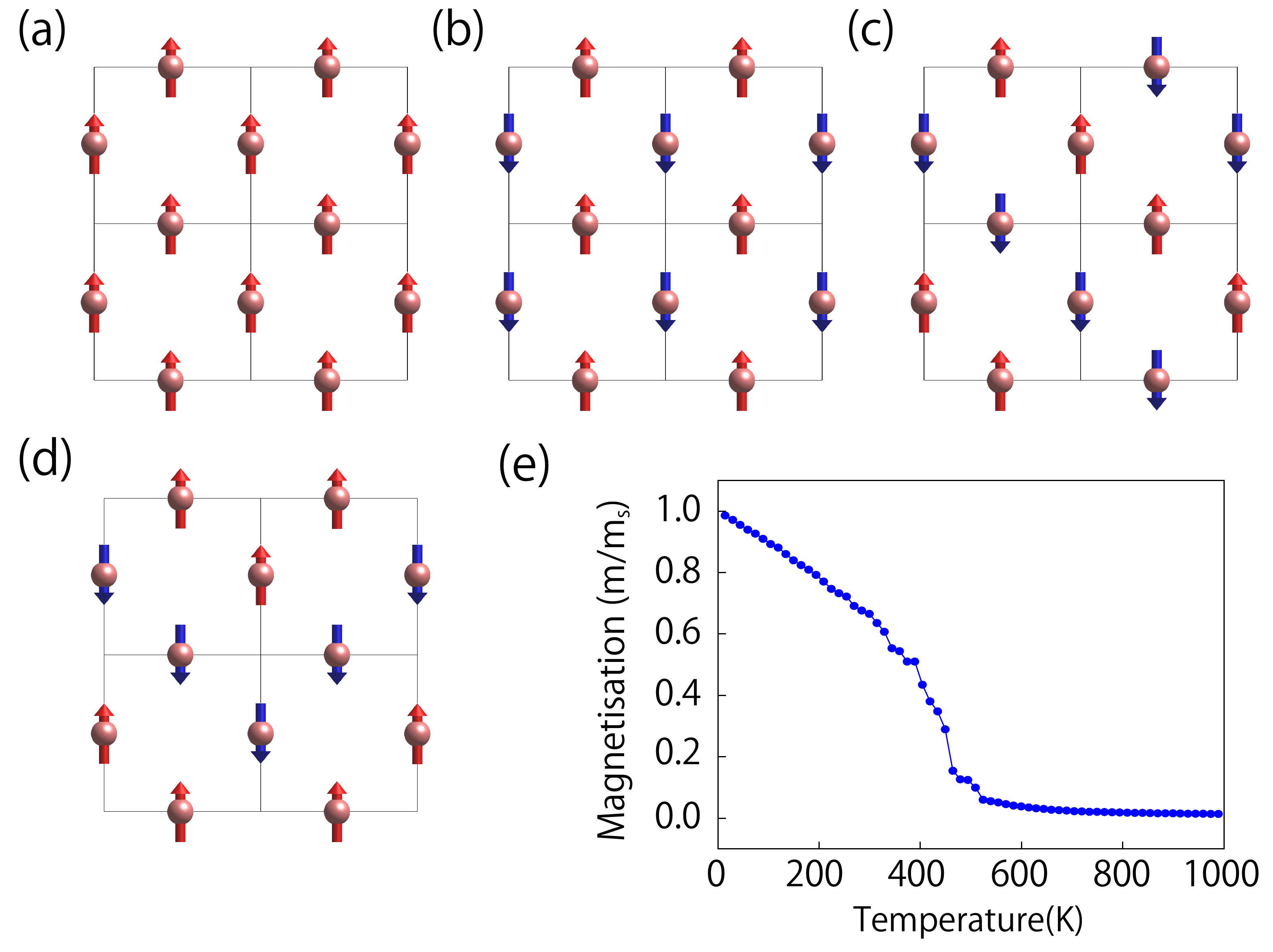}
	\caption{Schematic illustrations of the four magnetic configurations investigated: (a) ferromagnetism (FM), (b) altermagnetism (AM), (c) striped-type antiferromagnetism (SAFM), and (d) zigzag-type antiferromagnetism (ZAFM). (e) Temperature dependence of the staggered magnetization of monolayer Ti$_2$I$_2$S.}
	\label{fig2}
\end{figure}

The partially occupied $3d$ orbitals of Ti atoms give rise to intrinsic magnetic moments in these monolayers. To determine the magnetic ground state, four magnetic configurations were considered [see Fig.~\ref{fig2}(a)--(d)]: ferromagnetic (FM), altermagnetic (AM), stripe antiferromagnetic (stripe-AFM), and zigzag antiferromagnetic (zigzag-AFM). Total energy calculations confirm that the AM configuration is the energetically favored ground state for Ti$_2$I$_2$S. The corresponding energy differences, including those of other AM candidate materials, are summarized in Table~\ref{table1}. In this ground state, the magnetic moments are primarily localized on the Ti atoms, each carrying approximately $2~\mu_B$.
In the absence of SOC, the two spin sublattices are related by the $\{C_{2}\parallel C_{4z}^+\}$ symmetry operations, identifying these monolayers as altermagnetic materials. We further estimate the Néel temperature ($T_N$) of the AM ground state using Monte Carlo (MC) simulations based on an effective spin Hamiltonian~\cite{evans2014atomistic}:
\begin{equation}\label{Heisenberg}
	H=-\sum_{ i, j} J_{i j} \bm{S}_{i} \cdot \bm{S}_{j}-K\sum_{i}\left(S_{i}^{z}\right)^{2},
\end{equation}
where $\bm S_i$ is the normalized spin vector on the Ti site $i$, $J_{ij}$ is the exchange coupling constant between sites $i$ and $j$, and $K$ is the site anisotropy strength. All model parameters are extracted from first-principles calculations. The nearest-neighbor exchange interaction $J_1$ and the magnetic anisotropy constant $K$ are summarized in Table~\ref{table1}. The MC simulations were performed using a $50\,\mathrm{nm} \times 50\,\mathrm{nm}$ system to minimize finite-size effects. At each temperature, $20{,}000$ equilibration steps were performed, followed by $50{,}000$ MC steps for statistical averaging. For monolayer Ti$_2$I$_2$S, the corresponding model parameters are $J_1 = -1.057 \times 10^{-20}\,\mathrm{J}$ and $K = 1.15 \times 10^{-23}\,\mathrm{J}$.
The Néel temperature is determined from the temperature dependence of the sublattice magnetization, as shown in Fig.~\ref{fig2}(e). Based on the MC simulations, the Néel temperature of Ti$_2$I$_2$S is estimated to be approximately 450~K, which is well above room temperature and suggests the potential for realizing robust antiferromagnetic order at experimentally relevant temperatures. It should be emphasized, however, that this value is a theoretical estimate obtained from a parameterized spin Hamiltonian for an ideal monolayer. In realistic 2D samples, magnetic anisotropy, defects, and substrate interactions may modify the magnetic exchange interactions and anisotropy, thereby shifting the experimentally observed ordering temperature. Therefore, the calculated $T_N$ should be regarded as an estimate of the intrinsic magnetic ordering scale rather than a direct prediction of the experimental $T_N$.

\subsection{Electronic band and valley structures}
\begin{figure*}[htb]
	\includegraphics[width=15cm]{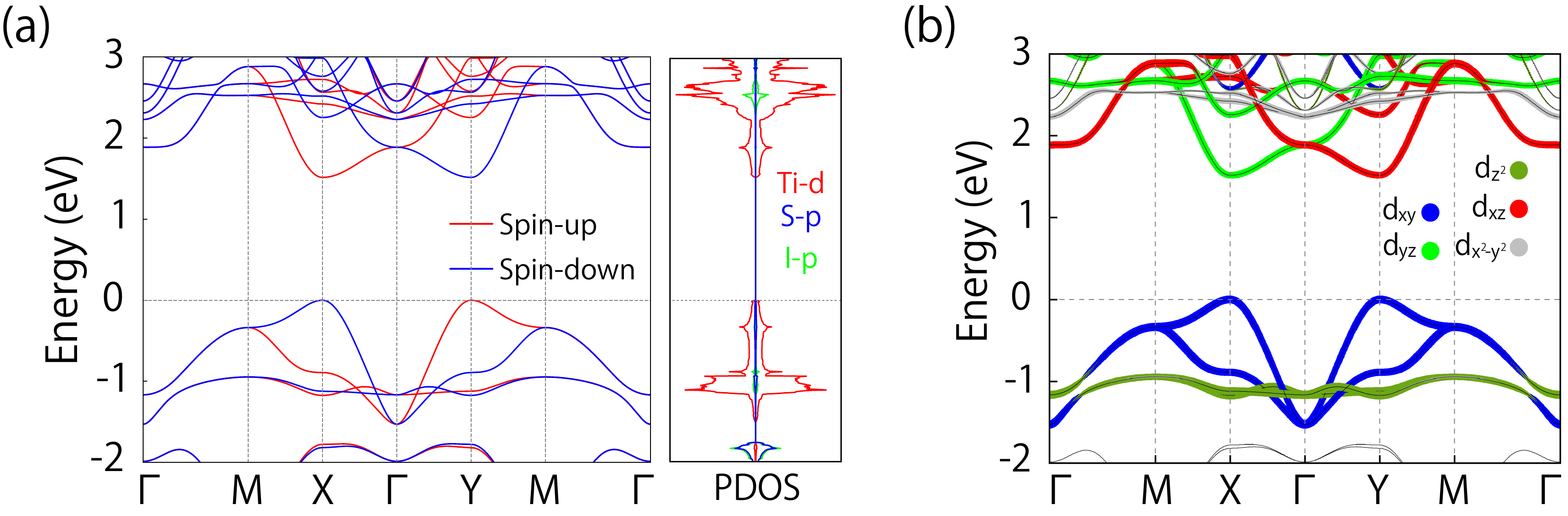}
	\caption{(a) Electronic band structures and projected density of states of monolayer Ti$_2$I$_2$S. (b) Orbital-resolved band structures projected onto the Ti atoms for monolayer Ti$_2$I$_2$S.}
	\label{fig3}
\end{figure*}
To begin with, we investigate the electronic band structure of monolayer Ti$_2$I$_2$S in the absence of SOC. The calculated spin-resolved band structures, projected density of states (PDOS), and orbital-resolved band structures are presented in Fig.~\ref{fig3}. As shown in the spin-resolved band structures in Fig.~\ref{fig3}(a), monolayer Ti$_2$I$_2$S exhibits a direct band gap of 1.515 eV alongside a pronounced spin splitting. Moreover, the spin splitting displays a characteristic $d$-wave pattern: the splitting changes sign between the $\Gamma$--X and $\Gamma$--Y directions, as dictated by the $\{C_{2}\parallel C_{4z}^+\}$ symmetry. Such momentum-dependent spin splitting in the absence of SOC is a hallmark of altermagnetism. Both the conduction band minimum (CBM) and the valence band maximum (VBM) are located at the high-symmetry points X and Y, manifesting distinct valley degrees of freedom. Notably, the spin splitting at the valence band edges at X and Y reaches as large as 0.892 eV. Despite their energy degeneracy, these valleys host opposite spin polarizations with completely separated spin-up and spin-down bands, thereby establishing a robust spin--valley locking. This behavior is fundamentally dictated by the underlying $\{C_{2}\parallel C_{4z}^+\}$ symmetry.
Furthermore, the PDOS and orbital-resolved band structures reveal that the states near the Fermi level are dominated by Ti $d$ orbitals, specifically originating from the $d_{xy}$ orbital for the valence bands and the $d_{yz}$ and $d_{xz}$ orbitals for the conduction bands. In addition, all bands along the $\Gamma$–M path remain spin-degenerate, a feature protected by the $\{C_{2}\parallel M_{110}\}$ symmetry.

\subsection{Strain-induced valley splitting and piezomagnetism}
\begin{figure*}[htb]
	\includegraphics[width=17cm]{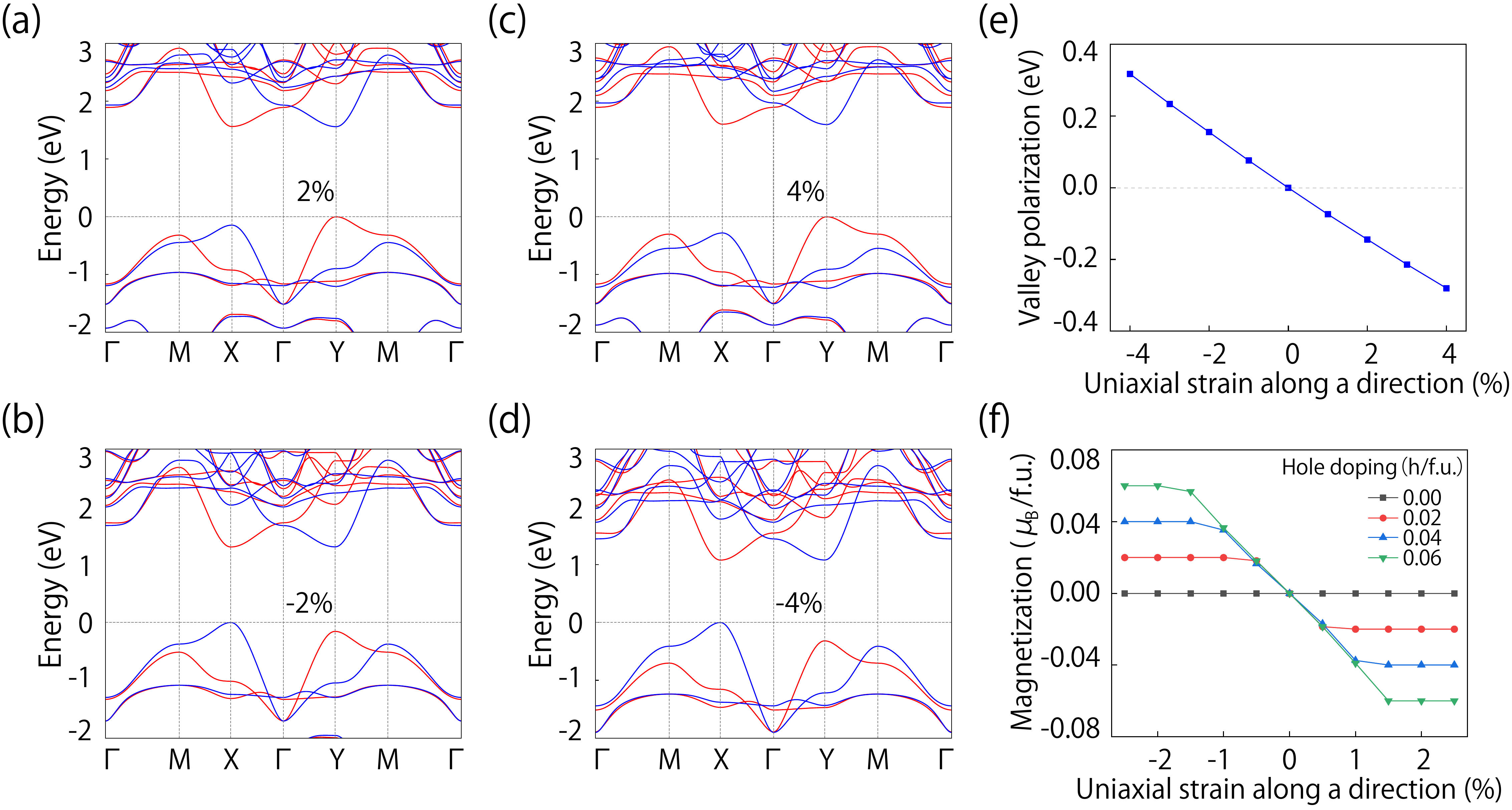}
	\caption{(a)–(d) Evolution of band structures for monolayer Ti$_2$I$_2$S under different uniaxial strains along the $a$-axis.  (e) Strain-induced valley polarization at the top of the valence band for monolayer Ti$_2$I$_2$S. (f) Net magnetization under different uniaxial strains along the $a$-axis with various hole doping levels.}
	\label{fig4}
\end{figure*}
As discussed above, the energetic degeneracy of the valleys at the X and Y points is strictly protected by the $\{C_{2}\parallel C_{4z}^+\}$ symmetry. Applying uniaxial strain will break this symmetry and lifts the degeneracy between the two valleys. As shown in Figs.~\ref{fig4}(a)--(d), tensile strain shifts the valence band at the Y point upward and that at the X point downward, thereby generating valley energy polarization. Conversely, compressive strain reverses these band shifts, with the valence band at Y moving downward and that at X upward, again leading to valley energy polarization.
To quantify this effect, we define the valley energy polarization as the energy difference $P = E(X) - E(Y)$ between the two valleys. The evolution of $P$ as a function of strain is presented in Fig.~\ref{fig4}(e). A giant strain-induced valley splitting is clearly observed. In particular, $P$ reaches 280 meV under a 4\% uniaxial strain, surpassing the valley splitting reported in representative 2D altermagnetic V$_2$Se$_2$O-family materials~\cite{ma2021multifunctional,li2024strain}.
Such strain-induced valley splitting provides an effective route to generate a net macroscopic magnetization in these 2D systems. Specifically, the net magnetic moment is given by $M = \int_{-\infty}^{E_f(n)} [\rho^\uparrow(\epsilon) - \rho^\downarrow(\epsilon)]\, d{\epsilon}$, where $E_f$ is the Fermi energy determined by the carrier density $n$, and $\rho^{\uparrow(\downarrow)}$ denotes the spin-resolved density of states.
As shown in Fig.~\ref{fig4}(f), Ti$_2$I$_2$S remains magnetically compensated under strain in the absence of doping. However, hole doping induces a finite magnetization that increases with the magnitude of the applied strain. From an experimental perspective, the required hole doping level of $0.015$ holes/f.u. corresponds to a moderate 2D carrier density of approximately $6 \times 10^{12}\ \text{cm}^{-2}$. Such carrier densities are readily accessible in modern 2D device architectures and can be continuously tuned using advanced electrostatic gating techniques, such as solid polymer electrolytes and high-dielectric-constant materials including hBN~\cite{deng2018gate,jiang2018controlling}, as well as through surface chemical doping. Notably, the sign of the magnetization reverses when switching between tensile and compressive strain. For a fixed strain, increasing the hole concentration further enhances the magnetic response. Moreover, the magnetization scales approximately linearly with strain at small deformations and gradually saturates at larger strain.

While uniaxial strain can effectively induce valley energy polarization, converting this energy splitting into an experimentally observable valley population imbalance requires appropriate thermodynamic and dynamical conditions. First, the carrier density must be properly controlled through chemical doping or electrostatic gating such that the Fermi level lies between the split valence-band edges of the X and Y valleys, thereby preferentially populating holes in the higher-energy valley. Second, the operating temperature should be sufficiently low to ensure that the thermal broadening remains much smaller than the strain-induced valley splitting. In Ti$_2$I$_2$S, $P$ can reach 280 meV under 4\% strain, which is more than an order of magnitude larger than the thermal energy at room temperature ($\sim 26$ meV), suggesting that a substantial valley population imbalance can remain robust against thermal fluctuations. Finally, appropriate relaxation dynamics are required: the intervalley scattering time should be sufficiently longer than the intravalley relaxation time to suppress thermal equilibration between the two valleys and preserve the steady-state population imbalance.

\subsection{Generation of noncollinear spin currents}

\begin{figure*}[htb]
	\includegraphics[width=15cm]{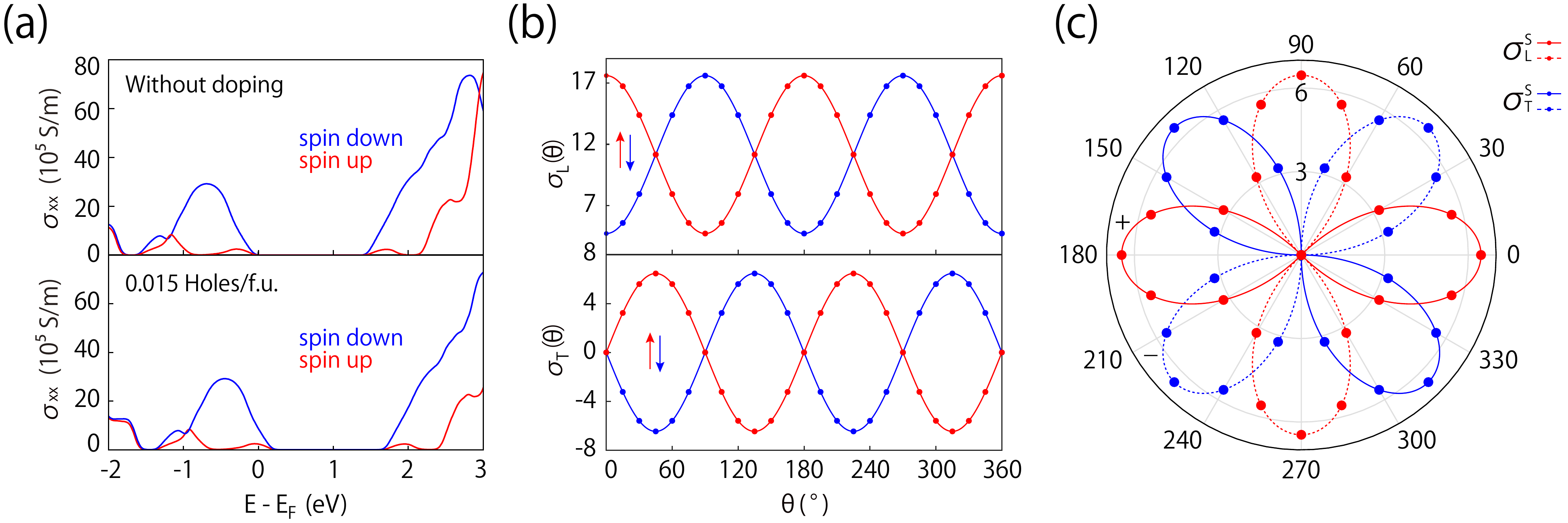}
	\caption{(a) Spin-resolved charge conductivity ($\sigma_{xx}$) of monolayer Ti$_2$I$_2$S without and with hole doping of 0.015 holes/f.u. (b) Angular dependence of the longitudinal ($L$) and transverse ($T$) charge conductivities as a function of the electric field direction angle $\theta$ for monolayer Ti$_2$I$_2$S. (c) Corresponding angular dependence of the longitudinal and transverse spin conductivities as a function of $\theta$. The “$+$” and “$-$” symbols denote positive and negative values, respectively.}
	\label{fig5}
\end{figure*}

When an in-plane electric field is applied, an altermagnetic system can generate a spin current. In the absence of SOC, spin remains a conserved quantum number, allowing the spin-up and spin-down charge conductivities ($\sigma_0^{\uparrow,\downarrow}$) to be evaluated independently. Figure~\ref{fig5}(a) shows the spin-resolved longitudinal conductivity $\sigma_{xx}$ of the Ti$_2$I$_2$S monolayer in its pristine state (top panel) and under hole doping of 0.015 per formula unit (bottom panel), with the electric field applied along the $x$ direction.
In the undoped case, $\sigma_{xx}$ vanishes at the Fermi level, confirming the insulating nature of the system. In contrast, hole doping induces a finite $\sigma_{xx}$ near the Fermi energy, rendering the system conductive. Owing to the anisotropic spin splitting of the electronic bands, a clear difference emerges between the spin-up and spin-down conductivities. The resulting spin current is characterized by the spin conductivity, defined as $\sigma_0^{S} = \sigma_0^{\uparrow} - \sigma_0^{\downarrow}$.
We further examine the angular dependence of the longitudinal ($\sigma_L^{\uparrow,\downarrow}$) and transverse ($\sigma_T^{\uparrow,\downarrow}$) conductivities with respect to the electric-field direction $\theta$. These components follow
$\sigma_{L}^{\uparrow,\downarrow}(\theta) = \sigma_0 \mp \sigma_0^{S}\cos 2\theta,\quad
\sigma_{T}^{\uparrow,\downarrow}(\theta) = \mp \sigma_0^{S}\sin 2\theta.
$
As shown in Fig.~\ref{fig5}(b), both conductivities exhibit a periodic dependence on $\theta$ with a period of $180^\circ$. The corresponding net spin conductivities are given by
$
\sigma_{L}^{S}(\theta) = \sigma_{L}^{\uparrow} - \sigma_{L}^{\downarrow} = -2\sigma_0^{S}\cos 2\theta,\quad
\sigma_{T}^{S}(\theta) = \sigma_{T}^{\uparrow} - \sigma_{T}^{\downarrow} = -2\sigma_0^{S}\sin 2\theta,
$
as plotted in Fig.~\ref{fig5}(c). Notably, a $45^\circ$ phase shift exists between the longitudinal and transverse components. At $\theta = 45^\circ$, the longitudinal spin conductivity $\sigma_{L}^{S}$ vanishes, yielding a purely transverse spin current.

\begin{figure*}[htb]
	\includegraphics[width=15cm]{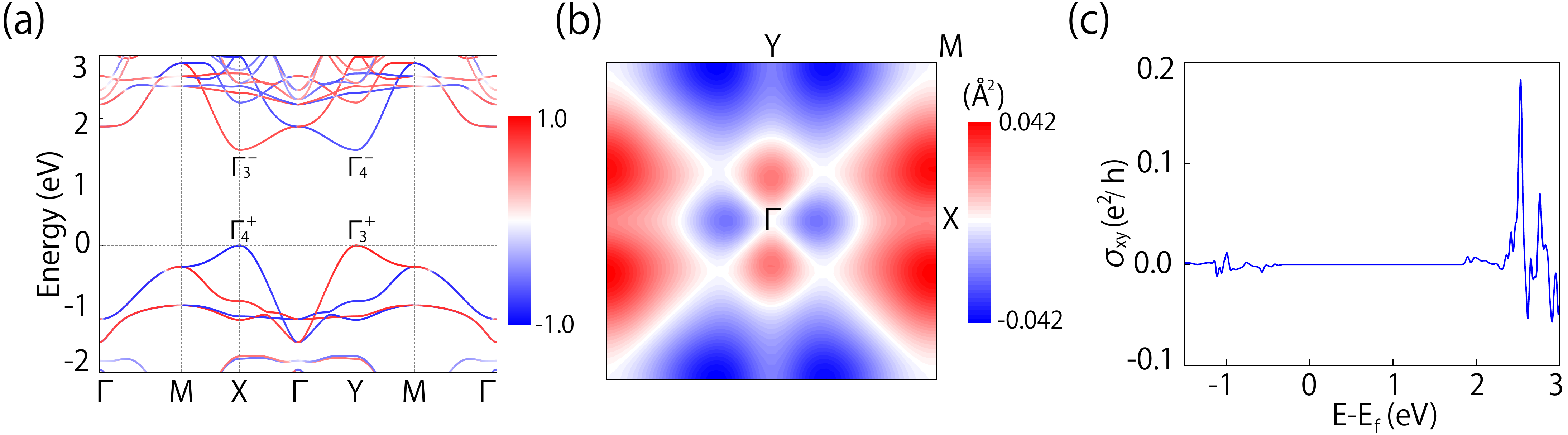}
	\caption{(a) Band structure of monolayer Ti$_2$I$_2$S including SOC, with the spin projection $s_z$ indicated, where the irreducible representations at the band edges are explicitly labeled. (b) Berry-curvature distribution summed over all valence bands. (c) Anomalous Hall conductivity of monolayer Ti$_2$I$_2$S.}
	\label{fig6}
\end{figure*}

{\subsection{Valley-contrasting Berry curvature and anomalous Hall conductivity}}
Another notable feature of 2D altermagnetic systems is their ability to host a nonzero Berry curvature and exhibit a finite anomalous Hall conductivity despite their fully compensated macroscopic magnetization. We first examine the orientation of the N\'{e}el vector and the electronic band structures of monolayer Ti$_2$I$_2$S with the inclusion of SOC. The magnetocrystalline anisotropy energy (MAE) is derived by setting the magnetization parallel to the [001], [100], and [110] crystallographic axes and evaluating the resulting total energies. Our calculations show that the magnetic easy axis of monolayer Ti$_2$I$_2$S is oriented along the [001] direction (see Table~\ref{table1}).
Figure~\ref{fig6}(a) presents the energy band structure of monolayer Ti$_2$I$_2$S with SOC included, where the red and blue colors indicate the spin projection, $\langle n\mathbf{k}|\hat{s}_z|n\mathbf{k}\rangle$. It is evident that SOC only slightly modifies the band structure, and the system retains its direct-gap semiconducting character, with the band gap reduced by only 6.9 meV compared with that without SOC.

We next turn to the Berry curvature and the anomalous Hall effect. In 2D systems, the Berry curvature possesses only an out-of-plane (z) component, which behaves as a pseudoscalar. For a given Bloch state $|n\bm k\rangle$, it is expressed as 
\begin{equation}\label{BC}
	\Omega_{n\bm k}=-2 \operatorname{Im} \sum_{n'\neq n} \frac{\left\langle n \bm{k}\left|v_{x}\right| n' \bm{k}\right\rangle\left\langle n' \bm{k}\left|v_{y}\right| n \bm{k}\right\rangle}{(\omega_{n^{\prime}}-\omega_{n})^{2}},
\end{equation}
in which $v_{x/y}$ stand for the velocity operators and $E_n=\hbar\omega_{n}$ denotes the energy eigenvalue of $|n\bm k\rangle$. The cumulative Berry curvature, obtained by summing over all occupied valence bands, $\Omega(\bm k)=\sum_{n\in occ.} \Omega_{n\bm k}$, is shown in Fig.~\ref{fig6}(b). Pronounced Berry-curvature hotspots with opposite signs are localized at the X and Y valleys, demonstrating a distinct valley-contrasting behavior. From a semiclassical perspective~\cite{xiao2010berry}, the Berry curvature acts as an effective magnetic field in momentum space, giving rise to an anomalous carrier velocity proportional to $\bm E\times\bm{\Omega}$ under an applied electric field. Consequently, the opposite Berry curvatures at the two valleys enable a valley Hall effect, allowing valley-resolved transverse currents to be generated by an in-plane electric field. Moreover, the nonvanishing Berry curvature also gives rise to an intrinsic anomalous Hall effect in the absence of an external magnetic field. To quantify this behavior, we calculate the intrinsic anomalous Hall conductivity, a macroscopic transport coefficient determined solely by the momentum-space geometry of the occupied electronic states. Within the first-principles formalism~\cite{jungwirth2002anomalous,yao2004first}, the intrinsic anomalous Hall conductivity is evaluated as
\begin{equation}
	\sigma_{x y}=-\frac{e^{2}}{\hbar} \int_\text{BZ} \frac{d^2 k}{(2\pi)^2} \Omega_z\left(\bm k\right).
\end{equation}
In this expression, $\Omega_z\left(\bm k\right)$ denotes the out-of-plane component of the cumulative Berry curvature, obtained by summing over all occupied bands at a given crystal momentum $\bm k$, as defined in Eq.~(\ref{BC}). 
Figure~\ref{fig6}(c) presents the calculated anomalous Hall conductivity, $\sigma_{x y}$, as a function of the chemical potential. As the chemical potential is tuned away from the band gap, $\sigma_{x y}$ becomes finite and exhibits pronounced peaks, with magnitudes comparable to those reported for conventional transition-metal ferromagnets.

\subsection{Linear dichroism}
\begin{figure*}[htb]
	\includegraphics[width=14cm]{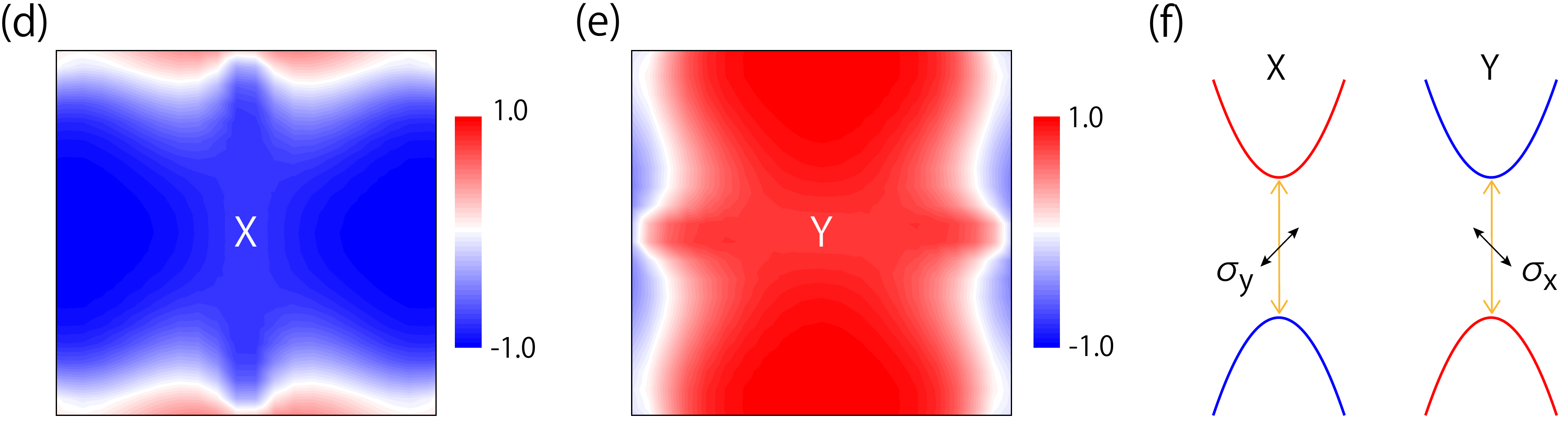}
	\caption{Linear dichroism $\xi(\bm{k})$ for monolayer Ti$_2$I$_2$S around the (a) X and (b) Y valleys, calculated using DFT with SOC included. (c) Schematic illustration of the optical transition selection rules at the two valleys, where red (blue) denotes spin-up (spin-down) bands.}
	\label{fig7}
\end{figure*}
Altermagnetic materials possessing valley degrees of freedom can exhibit valley-selective linear dichroism~\cite{wang2026two,li2026quantum}, whereby the polarization direction of incident linearly polarized light enables selective excitation of carriers in a particular valley. To quantify this effect, we define the $\bm{k}$-resolved linear dichroism parameter, $\xi(\bm{k})$, as
\begin{equation}
	\xi(\bm k) = \frac{|\mathcal{M}_x(\bm k)|^2-|\mathcal{M}_y(\bm k)|^2}{|\mathcal{M}_x(\bm k)|^2+|\mathcal{M}_y(\bm k)|^2},
\end{equation}
where $\mathcal{M}_i=\langle u_c(\bm{k})|\partial_{k_i}\mathcal{H}|u_v(\bm{k})\rangle$ ($i=x,y$) denotes the interband optical transition matrix element for linearly polarized light along the $i$ direction, $|u_{c(v)}(\bm{k})\rangle$ is the cell-periodic part of the Bloch wave function for the conduction (valence) band, and $\mathcal{H}$ is the Bloch Hamiltonian.
For monolayer Ti$_2$I$_2$S, the conduction- and valence-band states at the X point belong to the $\Gamma_3^{-}$ and $\Gamma_4^{+}$ irreducible representations of the $C_{2h}$ point group, respectively. By contrast, at the Y point they belong to the $\Gamma_4^{-}$ and $\Gamma_3^{+}$ representations [see Fig.~\ref{fig6}(a)]. Figures~\ref{fig7}(a) and \ref{fig7}(b) display the DFT-calculated $\xi(\bm{k})$ including SOC in the vicinity of the X and Y valleys. Notably, $\xi(\bm{k})$ is negative around the X valley but becomes positive around the Y valley. This sign reversal is dictated by the altermagnetic $C_{4z}\mathcal{T}$ symmetry. Consequently, optical transitions at the X (Y) valley couple predominantly to $y$- ($x$-) polarized light, demonstrating pronounced valley-selective linear dichroism, as schematically illustrated in Fig.~\ref{fig7}(c). Therefore, illumination with $x$- ($y$-) polarized light selectively excites carriers in the Y (X) valley, generating a valley-polarized carrier population. Such optically generated valley polarization provides a promising route toward future valleytronic applications.

\subsection{Magneto-optical Kerr effect}
\begin{figure*}[htb]
	\includegraphics[width=16cm]{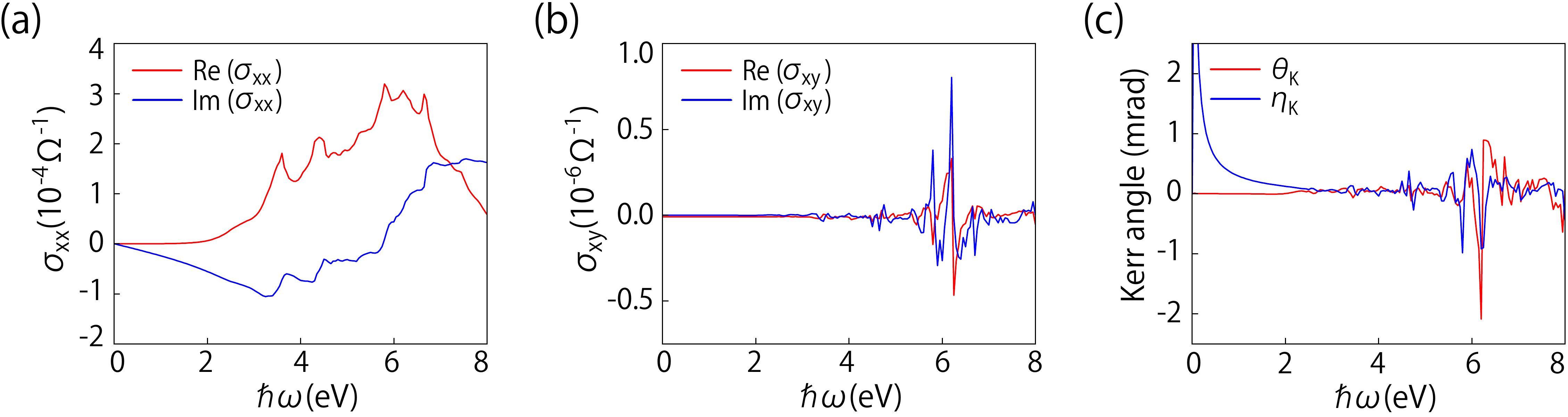}
	\caption{Calculated (a) diagonal ($\sigma_{xx}$) and (b) off-diagonal ($\sigma_{xy}$) components of the optical conductivity tensor for monolayer Ti$_2$I$_2$S. (c) Kerr rotation angles $\theta_K$ and Kerr ellipticities $\eta_K$ for monolayer Ti$_2$I$_2$S.}
	\label{fig8}
\end{figure*}
The magneto-optical Kerr effect (MOKE) serves as a powerful probe for investigating the electronic and magnetic properties of materials. For solids possessing at least threefold rotational symmetry, the optical conductivity tensor satisfies $\sigma_{xx}=\sigma_{yy}$ and $\sigma_{xy}=-\sigma_{yx}$. To evaluate the optical conductivity tensor $\boldsymbol{\sigma}$ quantitatively, we employ the Kubo--Greenwood formalism~\cite{yates2007spectral},
\begin{equation}\label{eq:OPC}
	\begin{aligned}
		\sigma_{\mu\nu} =\, & \frac{ie^2\hbar}{N_k V}\sum_{\bm{k}}\sum_{n,m}
		\frac{f_{m\bm{k}}-f_{n\bm{k}}}{E_{m\bm{k}}-E_{n\bm{k}}} \\
		&\times
		\frac{
			\langle\psi_{n\bm{k}}|\hat{\upsilon}_{\mu}|\psi_{m\bm{k}}\rangle
			\langle\psi_{m\bm{k}}|\hat{\upsilon}_{\nu}|\psi_{n\bm{k}}\rangle
		}{E_{m\bm{k}}-E_{n\bm{k}}-(\hbar\omega+i\eta)},
	\end{aligned}
\end{equation}
where $V$ is the unit-cell volume, $N_k$ is the total number of $\bm{k}$-points sampled in the BZ, $\omega$ is the incident photon frequency, and $\eta$ is a phenomenological broadening parameter. $f_{n\bm{k}}$ is the Fermi--Dirac distribution function, $\hat{\upsilon}_{\mu(\nu)}$ is the velocity operator along Cartesian direction $\mu,\nu\in\{x,y,z\}$, and $\psi_{n\bm{k}}$ and $E_{n\bm{k}}$ are the Wannier-interpolated Bloch state and its energy eigenvalue at band index $n$ and crystal momentum $\bm{k}$, respectively.
In the polar geometry, the complex Kerr angle for a material with at least threefold rotational symmetry takes the form
\begin{equation}
	\theta_K+i\eta_K
	=
	\frac{-\sigma_{xy}}{\sigma_{xx}\sqrt{1+i(4\pi/\omega)\sigma_{xx}}},
\end{equation}
where $\theta_K$ and $\eta_K$ are the Kerr rotation angle and Kerr ellipticity, respectively. Both quantities are directly accessible from the optical conductivity tensor computed from the material's electronic structure.
Figure~\ref{fig8} presents the calculated optical conductivity, Kerr rotation angles ($\theta_K$), and Kerr ellipticities ($\eta_K$) of monolayer Ti$_2$I$_2$S.
As shown in Fig.~\ref{fig8}(a), the real part of the diagonal conductivity, $\mathrm{Re}(\sigma_{xx})$, vanishes below $\sim$2.0 eV, reflecting the semiconducting optical band gap of the system, and exhibits multiple interband transition peaks in the 3.0–7.0 eV range. Concurrently, the imaginary part, $\mathrm{Im}(\sigma_{xx})$, shows the corresponding dispersive behavior governed by the Kramers–Kronig relations.
The off-diagonal components, $\mathrm{Re}(\sigma_{xy})$ and $\mathrm{Im}(\sigma_{xy})$ [Fig.~\ref{fig8}(b)], which determine the magneto-optical response, remain nearly zero at low energies but display pronounced oscillations between 5.5 and 7.0 eV. Consequently, the calculated polar MOKE spectra are shown in Fig.~\ref{fig8}(c). The Kerr rotation angle $\theta_K$ exhibits a sharp resonance peak around 6.3 eV, reaching a maximum absolute value of approximately 2.0 mrad. This strong magneto-optical response in the ultraviolet regime highlights the robust coupling between light and spin-polarized electronic states, indicating the potential of monolayer Ti$_2$I$_2$S for optospintronic applications.

\bigskip
\section{Conclusion}
In summary, we identify monolayer titanium-based chalcogenide halides, Ti$_2X_2Y$ ($X$ = F, Cl, Br, I; $Y$ = O, S, Se, Te), as a new family of two-dimensional altermagnetic valley materials.  These monolayers host a robust $d$-wave altermagnetic ground state with momentum-dependent spin splitting in the absence of SOC. Enabled by the unique $\{C_{2}\parallel C_{4z}^+\}$ crystalline symmetry, the X and Y valleys remain energetically degenerate while carrying opposite spin polarizations, providing an intrinsic degree of freedom for valley manipulation. We further demonstrate that uniaxial strain lifts the valley degeneracy, resulting in giant valley polarization accompanied by a pronounced piezomagnetic response. In addition, the coexistence of altermagnetism and valley physics gives rise to a rich spectrum of spin- and valley-dependent transport and optical phenomena, including electric-field-driven noncollinear spin currents, the anomalous Hall effect, linear dichroism, and the magneto-optical Kerr effect upon inclusion of SOC. Our work establishes Ti$_2$X$_2$Y monolayers as a versatile platform for investigating the interplay between altermagnetism and valley physics, broadens the materials landscape of two-dimensional altermagnetic valley systems, and opens new opportunities for the design of multifunctional spintronic, valleytronic, and magneto-optical devices.

\bigskip
\begin{acknowledgements}
	This work was supported by the Key Program of the Natural Science Basic Research Plan of Shaanxi Province (Grant No. 2025JC-QYCX-007) and the Youth Project (Category B) of the Natural Science Basic Research Plan of Shaanxi Province (Grant No. 2026JC-YXQN-026).
\end{acknowledgements}


%

\end{document}